\documentclass[aps,prd,superscriptaddress,twocolumn,showpacs]{revtex4}
\usepackage{amsfonts,amssymb,amscd,amsmath}
\usepackage{graphicx}
\usepackage{epsfig}
\usepackage{latexsym}
\usepackage{amssymb}
\usepackage{subfigure}
\usepackage{hyperref}

\begin{document}

\def\abs#1{ \left| #1 \right| }
\def\lg#1{ | #1 \rangle }
\def\rg#1{ \langle #1 | }
\def\lrg#1#2#3{ \langle #1 | #2 | #3 \rangle }
\def\lr#1{\langle #1 \rangle }
\def\me#1{ \langle #1 \rangle }

\newcommand{\bra}[1]{\left\langle #1 \right\vert}
\newcommand{\ket}[1]{\left\vert #1 \right\rangle}
\newcommand{\bx}{\begin{matrix}}
\newcommand{\ex}{\end{matrix}}
\newcommand{\be}{\begin{eqnarray}}
\newcommand{\ee}{\end{eqnarray}}
\newcommand{\nn}{\nonumber \\}
\newcommand{\no}{\nonumber}
\newcommand{\de}{\delta}
\newcommand{\lt}{\left\{}
\newcommand{\rt}{\right\}}
\newcommand{\lx}{\left(}
\newcommand{\rx}{\right)}
\newcommand{\lz}{\left[}
\newcommand{\rz}{\right]}
\newcommand{\inx}{\int d^4 x}
\newcommand{\pu}{\partial_{\mu}}
\newcommand{\pv}{\partial_{\nu}}
\newcommand{\au}{A_{\mu}}
\newcommand{\av}{A_{\nu}}
\newcommand{\p}{\partial}
\newcommand{\ts}{\times}
\newcommand{\ld}{\lambda}
\newcommand{\al}{\alpha}
\newcommand{\bt}{\beta}
\newcommand{\ga}{\gamma}
\newcommand{\si}{\sigma}
\newcommand{\ep}{\varepsilon}
\newcommand{\vp}{\varphi}
\newcommand{\zt}{\mathrm}
\newcommand{\bb}{\mathbf}
\newcommand{\dg}{\dagger}
\newcommand{\og}{\omega}
\newcommand{\Ld}{\Lambda}
\newcommand{\m}{\mathcal}
\newcommand{\dm}{{(k)}}

\title{Optimal Estimation of a Classical Force with a Damped Oscillator
 in the non-Markovian Bath}

\author{Yang Gao} \email{gaoyangchang@gmail.com}
\affiliation{Department of Physics, Xinyang  Normal University,
Xinyang, Henan 464000, China}

\author{Hwang Lee}
\affiliation{Hearne Institute for Theoretical Physics and
Department of Physics and Astronomy, \\
Louisiana State University, Baton Rouge, LA 70803, USA }

\author{Yong Lei Jia}
\affiliation{Department of Physics, Xinyang  Normal University,
Xinyang, Henan 464000, China}

\date{\today }

\begin{abstract}
We solve the optimal quantum limit of probing a classical force
exactly by a damped oscillator initially prepared in the factorized
squeezed state. The memory effects of the thermal bath on the
oscillator evolution are investigated. We show that the optimal
force sensitivity obtained by the quantum estimation theory
approaches to zero for the non-Markovian bath, whereas approaches to
a finite non-zero value for the Markovian bath as the energy of the
damped oscillator goes to infinity.
\end{abstract}

\pacs{03.65.Ta, 06.20.Dk, 42.50.Dv, 42.50.St}
\maketitle

\section{Introduction}

The precisions of recent experiments detecting tiny forces and
displacements have reached so extremely high levels that the quantum
limits become important \cite{nanoligo,exper}. Many useful bounds
for ideal systems have been obtained \cite{glm}. However, for these
extremely sensitive measurements, the unavoidable bath-induced noise
must be taken into account \cite{open}. The problem of finding the
ultimate quantum limit in the open system is usually difficult.
There are only several exceptions that can be solved rigorously
\cite{excep}. While the estimation of a single parameter in the
Markovian bath has been extensively studied theoretically and
experimentally \cite{glm,exper}, the estimation in the non-Markovian
bath has not been received enough attention (notable contributions
are Refs. \cite{nmb}). In this paper, we take a further step forward
in this direction and address the estimation of the amplitude of a
classical force with known waveform probed by a damped
quantum-mechanical oscillator (the estimation of the waveform of a
force acting on an ideal harmonic oscillator was considered in Ref.
\cite{waveform}), and especially investigate the effects of
non-Markovianity of bath on the force sensitivity.

The damped oscillator could be a trapped ion under an external
electric field, or a mesoscopic mechanical slab submitted to a weak
force, or yet an end mirror driven by some gravitational wave in an
optical interferometer. Therefore, the precise detection of the
amplitude of a classical force plays an important role in the
domains of nanophysics and gravitational waves \cite{nanoligo}. This
problem was pioneered by the works of Braginsky and collaborators
\cite{brag} and Caves {\it et al.} \cite{bk}. It is an example of
the general problem of quantum estimation theory \cite{qet}. A
typical parameter estimation consists in sending a probe in a
suitable initial state through some parameter-dependent physical
channel and measuring the final state of the probe. Let $x$ be the
parameter to be estimated, $y$ be the outcome of the measurement,
and $X(y)$ be the estimator of $x$ constructed from the outcome $y$.
To quantify the quality of this estimation, a local parameter
sensitivity of $x$ is defined as $\delta^2 x = \int dy (X(y)-x)^2
p(y|x)$, where $p(y|x)$ is the conditional probability distribution
of obtained a certain outcome $y$ given $x$. The maximization of
$\delta^2 x$ over all possible measurement procedures leads to the
so-called quantum Fisher information (QFI) \cite{uncert}. It is
shown in Ref. \cite{uncert} that the QFI can be computed from
Uhlmann's quantum fidelity between two outgoing final states
corresponding to two different parameters.

In general, the analytical expression for QFI in the presence of
thermal noise is formidable, except for very particular situations
\cite{excep,latu}. Under the Markovian approximation, Latune {\it et
al.} in Ref. \cite{latu} have obtained a tight bound for the force
sensitivity by applying a variational method proposed in Refs.
\cite{varia} and a properly selected homodyne measurement. This
bound approaches to a finite but nonzero value, known as the
``potential sensitivity" in Ref. \cite{brag}, as the mean energy of
the damped oscillator goes to infinity. On the other hand, as the
memory of bath becomes important, the results in Ref. \cite{latu}
should be updated. It is necessary to consider the effects of
non-Markovianity of bath on the estimation of a classical force.
Through exact solution and numerics, we demonstrate that the optimal
force sensitivity in a non-Markovian bath can surpass the
``potential sensitivity". That is to say, the force sensitivity with
a sequence of discrete measurements can attain zero, as the mean
energy of the damped oscillator goes to infinity.

This paper is organized as follows. Section II summarizes the main
results of the exact dynamics of damped harmonic oscillator in a
general non-Markovian bath. The evolution from a pure squeezed state
is obtained through the Wigner characteristic function. In Section
III we review the Markovian dynamics of the damped oscillator. Then
in Section IV, we calculate the force sensitivities in terms of the
explicit expressions for QFI. Section V contains our numerical
results and theoretical analysis of the general behavior of QFI in
non-Markovian baths. Finally, a short conclusion is drawn in Section
VI.

\section{Exact dynamics of damped oscillator}

The standard microscopic model for the damped harmonic oscillator in
a general non-Markovian bath is based on the Hamiltonian $H$ of a
larger coupled oscillator-bath system \cite{fact}, \be H &=&
H_S+H_B+H_{int}\nn &=& \hbar \og_0 \lz {1 \over 2 }(p^2+x^2) - f
\varsigma(t) x \rz \nn && + \sum_n \lx {p_n^2 \over 2m_n}+ {1 \over
2 } m_n \omega_n^2 q_n^2\rx \nn && -q \sum_n c_n q_n+q^2 \sum_n
{c_n^2 \over 2m_n\og_n^2}. \label{hami} \ee Here the dimensionless
variables $p=P/\sqrt{m\hbar\og_0}$, $x=X\sqrt{m\og_0/\hbar}$, so
that $[q,p]=i$, and $f=F/\sqrt{m\hbar\og_0^3}$ are introduced in
terms of the mass $m$, the oscillation frequency $\og_0$, and the
force amplitude $F$. The oscillator is submitted to the classical
force $F\varsigma(t)$, and for simplicity, we assume that the time
variation of the force, $\varsigma(t)$, is already known, and such
that $\mathrm{max} |\varsigma(t)|=1$. Therefore, our aim is to give
the estimation of $f$. In the following, we choose $\og_0=1$ and the
natural units $\hbar=k_B=1$. It amounts to rescale energy in units
of $\hbar \og_0$.

Under the factorized initial condition $\rho(0)=\rho_S(0) \otimes
\rho_B$ with $\rho_B$ the Gibbs state of the bath alone at
temperature $T$, the reduced dynamics of the oscillator can be
rigorously obtained from the time-local Hu-Paz-Zhang (HPZ) master
equation \cite{fact}. It is also known that the HPZ master equation
is equivalent to the Heisenberg-Langevin (HL) equation under the
same initial condition.

In this paper, we use the resulting HL equation from (\ref{hami})
for the damped oscillator, i.e. \be \ddot x(t)+\int _{0} ^t
dt'\ga(t-t')\dot x(t')+ x(t) \nn = -\ga(t)x(0) +
F_T(t)+f\varsigma(t), \label{qle} \ee for further analysis. Here the
coupling with the thermal bath is described by an operator-valued
random force $F_T(t)$ and a mean force characterized by a memory
kernel $\ga(t)$. They satisfy the commutation relation,
$[F_T(t),F_T(0)]=i \dot \ga(t)$. In terms of the spectral density
defined as $J(\og)=\sum_n c_n^2\delta(\og-\og_n)/m_n\og_n$, we have
$\ga(t)=\int_0^\infty d\og J(\og) \cos (\og t)/\og$ and the
autocorrelation of $F_T(t)$ expressed by $\nu(t-t')\equiv
\lr{F_T(t)F_T(t')+F_T(t')F_T(t)}/2=\int_0^\infty d\og
J(\og)/2\coth(\og/2T) \cos \og (t-t')$. Integration of Eq.
(\ref{qle}) yields \be x(t)=X_q(t)+X_c(t), \label{sol} \ee where \be
X_q(t)&=& \dot G(t)x(0)+ G(t){\dot x(0)}+\int_0^tdt' G(t-t')F_T(t'),
\nn X_c(t)&=&\int_0^tdt' G(t-t')f\varsigma(t'). \ee Here the
retarded Green function $G(t)$ is the unique solution of $ \ddot
G(t)+\int_0^t dt' {\ga}(t-t')\dot G(t')+G(t)=0$ with the initial
conditions $G(0)=0$ and $\dot G(0)=1$.

To determine the final state of the forced oscillator, it is simple
to use the reduced Wigner characteristic function defined as
$\chi_t(P,Q)= \mathrm {Tr} \lz e^{-i(x(t)P+p(t)Q)} \rho(0) \rz$,
where $p(t)\equiv \dot x(t)$ is the momentum operator. From Eq.
(\ref{sol}) and the quadratic structure of $\rho_B$, we can describe
the final state with \be \chi_t=\chi_0(\dot {\m G},\m G) \exp \lz
-{1\over 2} \lx \beta_{x} P^2+\dot \beta_{x}PQ+ \beta_{p}
Q^2\rx-i\Phi \rz, \label{wign} \ee where $\chi_0=\mathrm {Tr} \lz
e^{-i(x(0)P+p(0)Q)} \rho_S(0) \rz$. As shown in Ref. \cite{latu},
all the quantities in the above equations depend on the spectral
density throughout $G(t)$, namely $\m G(t)=G(t)P+\dot G(t) Q$,
$\Phi(t)= X_c(t) P+\dot X_c(t) Q$, $\beta_x(t) = \int_0^t \int_0^t
dt_1 dt_2 G(t_1)G(t_2)\nu(t_1-t_2)$, and $\beta_p(t) = \int_0^t
\int_0^t dt_1 dt_2 \dot G(t_1)\dot G(t_2)\nu(t_1-t_2)$. It is worthy
to mention that these integrals usually admit no explicit analytical
expressions. From Eq. (\ref{wign}), we can see that the
contributions of the classical force are completely represented by
the phase term $\Phi(t)$.

Initially, we set the oscillator in the pure squeezed state
\cite{book}, $ \rho_S(0)=\lg \varepsilon \rg \varepsilon$ and $\lg
\varepsilon=S(\varepsilon)\lg 0$. Here the squeeze operator
$S(\varepsilon)=\exp\lz\lx\varepsilon a^{\dg\!\ 2}-\varepsilon^*
a^2\rx/2 \rz$ with $\varepsilon=re^{2i\theta}$ is applied on the
vacuum $\lg 0$ of the annihilation operators $a=(x+ip)/\sqrt{2}$.
The mean displacement and covariance matrix that fully describe such
a Gaussian state are $ \lr{a}=0$ and \be \Sigma_0 &=& \bigg( \bx
\lr{\tilde{a}^2} & \frac{1}{2}\lr{\tilde{a}\tilde{a}^\dg+
\tilde{a}^\dg \tilde{a}}\\
\frac{1}{2}\lr{\tilde{a}\tilde{a}^\dg+\tilde{a}^\dg \tilde{a}} &
\lr{\tilde{a}^{\dg 2}} \ex \bigg) \nn &=& \frac{1}{2} \bigg( \bx
e^{2i\theta}\sinh2r & \cosh 2r \\
\cosh 2r & e^{-2i\theta}\sinh 2r \ex \bigg), \ee where
$\tilde{a}=a-\lr{a}$ is the displaced operator. We point out that a
more physical initial state should consider the initial
oscillator-bath correlation \cite{pre}, such as the one prepared by
a projective measurement on the Gibbs state of the total system, $
\rho(0) \propto \lg \varepsilon \rg \varepsilon \otimes {\rg
\varepsilon e^{-H/T} \lg \varepsilon}$. For our cases, it has been
shown in Ref. \cite{gauss} that these two initial states lead to
almost the same results.

For the initial squeezed state, the final state is still a Gaussian
state characterized by \be \chi_t=\exp\lz -\frac{1}{2}\lx
\sigma_{xx} P^2+2\sigma_{xp} PQ+ \sigma_{pp} Q^2 \rx-i\Phi\rz,
\label{wigf} \ee where the coefficients are \be \sigma_{xx}(t) &=&
{G_x^2\over 2 \xi} +{G_p^2\xi \over 2}+\beta_{x}, \nn
\sigma_{pp}(t)&=& {\dot G_x^2\over 2\xi} +{\dot G_p^2\xi \over
2}+\beta_{p}, \label{fff} \ee and $\sigma_{xp} =\dot \sigma_{xx}/2$
with the notations $G_x=G\cos\theta -\dot G\sin\theta$,
$G_p=G\sin\theta+\dot G\cos\theta$, and $\xi=e^{2r}$. For an ideal
oscillator with $J(\og)=0$, we have $G=\sin t$ and
$\beta_x=\beta_p=0$.

\section{The Markovian dynamics}

It is known in Ref. \cite{book} that the effect of a Markovian bath
of mode $a_{\rm in}$ on a single mode boson $a$ can be described by
the following HL equation, \be \dot a=i
[H_S,a]-\frac{\ga}{2}a+\sqrt{\ga}a_{\rm in}, \label{hle} \ee where
$\ga$ is the damping strength. The bath mode $a_{\rm in}$ satisfies
the commutation relation $[a_{\rm in}(t'), a_{\rm
in}^\dg(t)]=\delta(t'-t)$, and $\lr{a_{\rm in}^\dg(t') a_{\rm
in}(t)}=n_T \delta(t'-t)$, where $n_T=(e^{1/T}-1)^{-1}$ is the
average excitation number at temperature $T$. The Markovian
evolution governed by the current form of HL equation is always
completely positive, as the corresponding master equation takes the
standard Lindblad form. The solution of Eq. (\ref{hle}) is \be
\tilde{a}(t)=g(t)a(0)+\sqrt{\ga}\int_0^t dt' g(t-t') a_{\rm in}(t'),
\label{msol} \ee where $g(t)=e^{-it-\ga t/2}$ and
$\tilde{a}=a(t)-\lr{a(t)}$ associated with $\lr
{a(t)}=\frac{1}{\sqrt{2}} [\lr {x(t)}+i\lr
{p(t)}]=\frac{i}{\sqrt{2}}\int_0^t dt' g(t-t')f\varsigma(t')$. From
Eq. (\ref{msol}) and the initial squeezed state, the final state is
characterized by $\lr{a(t)}$ and \be \Sigma_t = \frac{1}{2} \bigg(
\bx
d & c \\
c & d^* \ex \bigg),\ee where $c=e^{-\ga t}\cosh 2r + (1-e^{-\ga
t})(2N_T+1)$ and $d\equiv d_R+id_I=e^{2i(\theta-t)-\ga t}\sinh2r$.
Then the corresponding Wigner characteristic function is of the form
\be \chi_t=\exp \lz -{1\over 2} \lx \sigma_{xx} P^2+ 2\sigma_{xp}
PQ+ \sigma_{pp} Q^2\rx-i \Phi \rz, \ee where the
covariance matrix is \be \sigma \equiv \bigg( \bx \si_{xx} & \si_{xp} \\
\si_{xp} & \si_{pp} \ex \bigg) = \frac{1}{2} \bigg( \bx
c+d_R & d_I \\
d_I & c-d_R \ex \bigg), \ee and the phase is $\Phi=\lr {x} P+\lr {p}
Q$.

\section{QFI for force estimation}

The results obtained in the last two sections allow us to get the
QFI for probing a classical force with the damped oscillator.
According to the quantum estimation theory \cite{uncert}, the force
sensitivity is bounded by $\de^2 f \ge 1/\mathcal{H}$, where $\m H =
4(1-\mathcal{F}^2(\rho^S_{f+df},\rho^S_f))/d f^2$ is the QFI in
terms of the quantum fidelity
$\mathcal{F}(\rho_{2},\rho_1)=\mathrm{Tr}\sqrt{\sqrt{\rho_1} \rho_2
\sqrt{\rho_1}}$. The fidelity between two arbitrary states is
usually difficult to calculate analytically. However, for two
arbitrary Gaussian states $\rho_1$ an $\rho_2$ with respective
covariance matrix $\bb \si_{1,2}$ and mean displacement vector $\lr
{\bb x}_{1,2}$ with $\bb x^T=(x,p)$, an explicit expression of
$\mathcal{F}$ is given by \cite{fide}, \be
\mathcal{F}(\rho_1,\rho_2)=\frac{\exp\lx -\frac{1}{2}\bb {u}^T ({\bf
\si}_1+ {\bf \si}_2)^{-1} \bb
{u}\rx}{\sqrt{\Gamma+4\Pi}-\sqrt{4\Pi}}, \ee where $\Gamma=\det(\bb
\si_1+\bb \si_2)$, $\Pi=(\det \bb \si_1-1/4)(\det \bb \si_2-1/4)$,
and $\bb u = \lr {\bb x}_2- \lr {\bb x}_1$. For our cases, ${\bf
\si}_1={\bf\si}_2={\bf\si}$ and $\lr {\bb x}_f=f \bb b \equiv
f(b_x,b_p)^T$, we get $\m H={\bb{b}^T \mathbf{\si}^{-1} \bb {b}}$,
where \be \bb b(t;\tau)=\int_{0}^{\tau} dt' \varsigma(t+\tau-t')\lx
G(t'),\dot G(t')\rx^T , \no\ee for the non-Markovian bath, and \be
\bb b(t;\tau)=\int_0^\tau dt'\varsigma(t+\tau-t') e^{-\ga t'/2} \lx
\sin(t'),\cos(t') \rx^T, \no \ee for the Markovian bath.

The above lower bound is actually achievable with a time variant
homodyne measurement as $M(t)=\bb x^T \bb \sigma^{-1} \bb
b/\sqrt{\bb{b}^T \mathbf{\si}^{-2} \bb b}$. It can be seen that the
corresponding force sensitivity is just $ \de^2 f={\Delta^2 M/\abs{d
\lr M / d f}}^2=1/(\bb b^T \bb \sigma^{-1} \bb b)$, where $\Delta^2
M=\lr{M^2}-\lr{M}^2$ is the variance of $M$. On the other hand, to
get a better sensitivity, we demand optimizing $\m H$ over all
adjustable parameters, such as the rotation angle $\theta$ in
$S(\varepsilon)$.

Suppose a given mean energy of the oscillator as $E=({\xi
}+{\xi^{-1}})/4 \ge {1/2}$. With an ideal oscillator and choosing
the optimal angle $\theta=t-\arctan(b_x/b_p)$, the maximum of $\m H$
is given by $\m H = 2\xi (b_x^2+b_p^2)$ in terms of $b_x=\int_0^tdt'
\sin (t-t')\varsigma(t')$ and $b_p=\dot b_x$. As $E \to \infty$, $\m
H \approx 8(b_x^2+b_p^2) E$, which is known as the Heisenberg limit
\cite{fhl}.

For the Markovian bath, because both of the eigenvalues of $\bb
\sigma$ are independent of $\theta$ as the ideal oscillator, we can
always choose $\theta=t-\arctan(b_x/b_p)$ in such a way that $\bb b$
becomes the eigenvector of $\bb \sigma$ associated with the minimal
eigenvalue $\ld_{\mathrm{min}}=1/2[(1-e^{-\ga t})(2n_T+1)+e^{-\ga
t}/\xi]$. This yields the maximum of the QFI, $\m H_{\mathrm{max}} =
|\bb b|^2/\ld_{\mathrm{min}}$, which is the same as Eq. (31) in Ref.
\cite{latu} obtained by a variational method proposed in Ref.
\cite{varia} and by a properly selected homodyne measurement.

\section{General behavior of QFI in the non-Markovian bath}

For the non-Markovian bath, the eigenvalues of $\bb \sigma$ are
complicated functions of $\theta$, and so is the optimal $\theta$
maximizing $\m H$. In general, it is impossible to represent the
final expression of $\m H$ analytically. We resort to numerics and
take the regularized Ohmic damping as an example \cite{open}, namely
$J(\og)=2\ga\og \exp({-{\og^2/\Ld^2}})/\pi$ in terms of the damping
strength $\ga$ and a cutoff scale $\Lambda$. For numerical purpose,
let the total measurement time be $t_{\rm tot}=\pi/2$, and the force
shape be $\varsigma(t)=1$ (constant force) or $\varsigma(t)=\cos t$
(resonant force). Other chosen parameters are $T=0$, $\ga=0.1$, and
$\Lambda=10$.

\begin{figure}[t!]\label{f1}
 \centering \subfigure[]{ \label{Fig.sub.a}
\includegraphics[width=0.47\columnwidth]{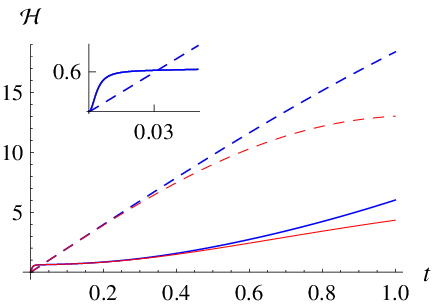}}\hspace{0.05in}
 \subfigure[]{ \label{Fig.sub.b}
\includegraphics[width=0.47\columnwidth]{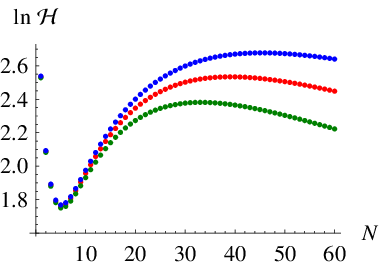}}\hspace{0.05in}
\caption{(a) The optimal $\m H$ over $\theta$ as a function of time
$t$ for a single measurement ($N=1$) on the damped oscillator in the
non-Markovian (solid) and Markovian (dashed) baths. The thick (thin)
lines are for the constant (resonant) force here and in the
following plots. The inset shows the short time behavior of $\m H$.
Here $r=5$. (b) The behavior of $\m H$ as a function of $N$ for
detecting a constant force in the non-Markovian bath with $r=2.50$,
$2.66$, and $2.80$ from the bottom up. Here the optimization of $\m
H$ over $\theta_k$ is made. }
\end{figure}

The plots of the optimal $\m H$ over $\theta$ as a function of the
measurement time $t$ are shown in Fig. 1 (a). We see that the
results for the non-Markovian and Markovian baths show quite
different behaviors as functions of $t$, especially at short times.
As $t\to 0$, we notice that $\m H \to {2\xi t^2}$ for the
non-Markovian bath, and $\m H \to {t / \ga}$ for the Markovian bath.
On the other hand, as $t\to \infty$, both of them approach to the
steady value $\m H \to b_x^2/\si^\infty_{xx}+b_p^2/\si^\infty_{pp}$.
It suggests to us that for a total probing time $t_{\rm tot}$, the
force sensitivity can be further improved by $N$-repeated
measurements, where each part with the mean energy $E$ is probed for
a time interval $\tau=t_{\rm tot}/N$. For this sequential strategy,
the corresponding QFI is the sum of the QFI for each measurement
step, \be \m H &=& \sum_{k=0}^{N-1} \bb b^T(k\tau;\tau)\bb
\sigma^{-1}(\theta_k,\tau)\bb b(k\tau;\tau), \label{qfi} \ee which
should be optimized over the variables $\theta_k$, $N$.

Fig. 1 (b) displays the behavior of $\m H$ versus $N$ for detecting
a constant force acting on the oscillator in the non-Markovian bath
with $r=2.50$, $2.66$, and $2.80$ from the bottom up. It shows that
the number $N$ that maximizes $\m H$ depends on the value of $r$. We
denote this optimal number by $N_{\rm opt}$, and $\tau_{\rm
opt}=t_{\rm tot}/N_{\rm opt}$. For $r\approx 2.66$, there are two
global maxima, corresponding to the values $1$ and $39$ of $N$.

Fig. 2 (a) displays the behavior of $N_{\rm opt}$ versus $r$ for the
non-Markovian and Markovian baths, respectively. Fig. 2 (b) displays
the behavior of $\m H_{\rm opt}$ versus $r$. As shown in Fig. 2, we
notice that for the ideal oscillator, the sequential strategy makes
no improvement, and $\m H_{\rm opt}$ is linearly proportional to $E$
\cite{bk}. For the damped oscillator, a single measurement should be
used for low $E$, and the sequential measurements are preferred for
increasingly high $E$. It can be seen that the free oscillator
presents the best force sensitivity among others. The non-Markovian
result is worse than the Markovian one for low $E$, and becomes
better as $E$ increases.

From Fig. 2 and for a sufficiently high $E \approx e^{2r}/4$, the
optimal number $N_{\rm opt}$ can be well approximated by \be N_{\rm
opt} \approx \bigg\{ \bx d_0 \sqrt{E} & \quad \text
{for the non-Markovian bath}, \\
c_0 E^{1/3}& \quad \text {for the Markovian bath}, \ex \ee and the
corresponding optimal QFI can be fitted by \be \m H_{\rm opt}
\approx \bigg\{ \bx d_1\sqrt{E}, \\ {c_1/\ga}-c_2E^{-{2/3}}. \ex \ee
In terms of the total energy $E_{\rm tot}=N E \to \infty$ and for an
arbitrary $t_{\rm tot}\gg 1$, we have \be \m H_{\rm opt}/t_{\rm
tot}\approx \bigg\{ \bx d_1'(E_{\rm tot}/t_{\rm tot})^{1/3} \to \infty, \\
{c'_1/\ga}-c'_2(E_{\rm tot}/t_{\rm tot})^{-1/2} \to {c'_1/\ga}, \ex
\ee where $d_1'=d_1(d_0\pi^2/4)^{-{1/3}}$, $c_1'= 2 c_1 /\pi$ and
$c_2'= c_2\sqrt{8c_0/\pi^3}$. We find that the above coefficients
take: $d_0=5.60$ independent of the force shapes, $d_1=1.76$ for
constant force, and $d_1=0.88$ for resonant force; $c_0=0.64$,
$c_1=31.41$, and $c_2=47.81$ for constant force, $c_0=0.80$,
$c_1=15.71$, and $c_2=30.11$ for resonant force. The Markovian
results are in agreement with Eqs. (33) and (35) in Ref.
\cite{latu}.

\begin{figure}[t!]
\centering \subfigure[]{ \label{Fig.sub.a}
\includegraphics[width=0.47\columnwidth]{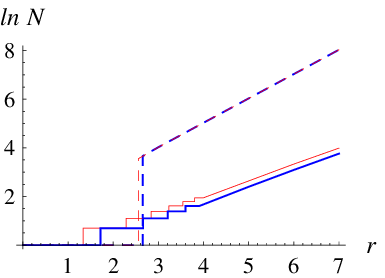}}\hspace{0.05in}
 \subfigure[]{ \label{Fig.sub.b}
\includegraphics[width=0.47\columnwidth]{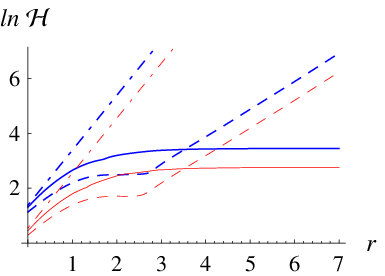}}\hspace{0.05in}
\caption{(a) The plots of $N_{\rm opt}$ versus the squeeze parameter
$r$ for the non-Markovian (dashed) and Markovian (solid)
oscillators. (b) The behavior of $\m H_{\rm opt}$ versus $r$ for the
free (dot-dashed), non-Markovian (dashed), and Markovian (solid)
oscillators. }
\end{figure}

The key point demonstrated by the above numerics is that the force
sensitivity with the damped oscillator initially prepared in a
squeezed state performs better for the non-Markovian case when the
total mean energy gets higher. Under such a situation, a relatively
larger $N_{\rm opt}$ is used and hence the oscillator can benefit
more from non-Markovian noise feature. Furthermore, the cube root
asymptotic for $\m H_{\rm opt}$ is not specific to our chosen model,
but rather a general consequence of the unitary evolution of the
total oscillator-bath state.

To put it more explicitly and following Ref. \cite{latu}, we
consider the limit of a fast sequential measurement, that is, $\tau$
is much smaller than all characteristic times of the process, and
are able to derive an analytical expression for $N_{\rm opt}$ and
$\m H_{\rm opt}$. Using the series expansions of $G(t)=t+G_3
t^3/3!+O(t^5)$ and $\nu(t)=\nu_0+\nu_2 t^2/2!+O(t^4)$, the optimal
angle is determined by $\theta_k=t/2+O(t^2)$ from the vanishing
derivative of Eq. (\ref{qfi}) with respect to $\theta_k$, and
therefore Eq. (\ref{qfi}) takes the form \be \m H = {2
\tau+O(\tau^3) \over 1/\xi+2 \nu_0 \tau^2+O(\tau^4)} \int_0^{t_{\rm
tot}}dt \varsigma^2(t), \label{nmb} \ee where the Euler-Maclaurin
formula has been applied to transform the summation over $k$ into an
integral, and therein the boundary terms have been chosen as zero.
Here the dominate term in the denominator starts from $O(t^2)$,
whereas for the Markovian case, it starts from $O(t)$ as shown in
Ref. \cite{latu}.

From the above equation, the optimal interval that maximizes $\m H$
is found to be $\tau_{\rm opt}=(2\nu_0\xi)^{-{1/2}}+O(\xi^{-1})$, or
\be N_{\rm opt}=t_{\rm tot}\sqrt{2\nu_0\xi}+O(1). \label{nopt}\ee
The optimal QFI is thus \be \m H_{\rm opt} = \sqrt{\xi \over
2\nu_0}\int_0^{t_{\rm tot}}dt \varsigma^2(t) +O(1). \label{hopt} \ee
The regime of validity of Eqs. (\ref{nopt}) and (\ref{hopt}) are
confined by the conditions $\tau_{\rm opt} \ll \mathrm {max}
\{1/\ga,1/\og_0,1/\Lambda,\tau_{\rm ch}\}$, where $\tau_{\rm ch}$ is
the characteristic time of evolution of $\varsigma(t)$. As $E \simeq
\xi/4 \to \infty$, Eq. (\ref{nopt}) implies that $ N_{\rm opt}$ is
independent of the force shape $\varsigma(t)$, while in Eq.
(\ref{hopt}) $\m H_{\rm opt}$ depends on $\varsigma(t)$ through the
integral $\int_0^{t_{\rm tot}}dt \varsigma^2(t)$, as confirmed by
the numerical results. For the regularized Ohmic bath at zero
temperature, we have $\nu_0=\ga\Lambda^2/(2\pi) \approx 1.59$,
$\int_0^{t_{\rm tot}}dt \varsigma^2(t) = t_{\rm tot}$ and $t_{\rm
tot}/2$ for the constant and resonant forces, respectively. They fit
the obtained numerical values quite well after crossing the point $r
\approx 2.66$.

\section{Conclusion}

In summary, we have found that the memory effects of the surrounding
bath could significantly change the short time behavior of the
system evolution. Using the quantum Fisher information, we have got
an exact expression for the optimal quantum limit of the force
sensitivity probed by the damped oscillator prepared in the
factorized initial squeezed state. The optimal force sensitivity
thus obtained approaches to zero for the non-Markovian bath, whereas
approaches to a finite non-zero value for the Markovian bath when
the mean energy of the oscillator goes to infinity.

\begin{acknowledgments}
The authors would like to think Profs. L. Davidovich and J. P.
Dowling for helpful discussions. This work is supported by NSFC
grand No. 11304265 and the Education Department of Henan Province
(No. 12B140013).
\end{acknowledgments}


\end{document}